\documentclass[11pt,latexsym,multicol,amsfonts,epsf,xy,amsbold,amsmath,amssymb,m
athsym]{article}

\usepackage[frame,line,arrow,matrix,tips]{xy}
\input{Qcircuit}

\def\C{{\mathbf{C}}}
\def\N{{\mathbf{N}}}

\date{October 20, 2006}

\newcommand{\cS}{{\cal S}}
\newcommand{\beq}{\begin{equation}}
\newcommand{\eeq}{\end{equation}}
\newcommand{\beqy}{\begin{eqnarray}}
\newcommand{\eeqy}{\end{eqnarray}}

\renewcommand{\>}{\rangle}

\newcommand{\polylog}{{\rm polylog}}
\newcommand{\poly}{{\rm poly}}

\newtheorem{Definition}{Definition}

\newenvironment{Definition*}{{\bf Definition}}{}

\title{Estimating Diagonal Entries of Powers of Sparse Symmetric Matrices 
is BQP-complete}

\author{Dominik Janzing\thanks{e-mail: janzing@ira.uka.de}\\ 
\small Institut f{\"u}r Algorithmen und Kognitive Systeme,
Universit{\"a}t Karlsruhe,\\ \small Am Fasanengarten 5,
D-76\,131 Karlsruhe, Germany\\
\\
Pawel Wocjan\thanks{e-mail: wocjan@cs.ucf.edu}\\
\small School of Electrical Engineering and Computer Science\\
\small University of Central Florida \\
\small Orlando, FL 32816, USA}

\begin{document}

\maketitle

\begin{abstract}
Let $A$ be a real symmetric matrix of size $N$ such that the number of the non-zero entries in each row is polylogarithmic in $N$ and the positions and the values of these entries are specified by an efficiently
computable function. We consider the problem of estimating an arbitrary diagonal entry $(A^m)_{jj}$ of
the matrix $A^m$ up to an error of $\epsilon\, b^m$, where $b$ is an a priori given upper bound on the norm of $A$, $m$ and $\epsilon$ are polylogarithmic and inverse polylogarithmic in $N$, respectively.

We show that this problem is BQP-complete. It can be solved efficiently on a quantum computer by repeatedly applying measurements of $A$ to the $j$th basis vector and raising the outcome to the $m$th power.  Conversely, every quantum circuit that solves a problem in BQP can be encoded into a sparse matrix such that some basis vector $|j\rangle$ corresponding to the input induces two different spectral measures depending on whether the input is accepted or not.  These measures can be distinguished by estimating the $m$th statistical moment for some appropriately chosen $m$, i.e., by the $j$th diagonal entry of $A^m$.  The problem is still in BQP when generalized to off-diagonal entries and it remains
BQP-hard  if $A$ has only $-1$, $0$, and $1$ as entries.
\end{abstract}

\section{Introduction}
It is believed that a quantum computer is more powerful than a classical computer in the sense that it makes possible to obtain faster algorithms for some computational problems than the best classical algorithms. 
However, it is still not understood well enough which problems are tractable for quantum computers.  It is therefore be desirable to better understand the class of problems which can be solved efficiently on a quantum computer. In quantum complexity theory, this class is referred to as BQP. Meanwhile, some characterizations of BQP are known \cite{KnillQuadr,PawelYard,aharonov-2006-,WZ:06}.  Here we present a new
characterization of BQP which is related to the computation of powers of large matrices.

It should not be too surprising that computational problems can be formulated in terms of ``large'' matrices.  For example, the transformations of a quantum computer can be represented by multiplication of matrices of a certain type.  However, the matrix problems derived from this representation, would usually not
be very natural in classical terms. They are, of course, natural, as physical questions about the behavior of quantum systems. For instance, the problem of estimating the entries of products of unitary matrices which are given by a tensor embedding of low-dimensional unitaries, is BQP-complete, but it is not obvious 
where problems of this nature could arise in real-life applications referring to the macroscopic world.

It is known that Hamiltonians with finite range interactions can generate sufficiently complex dynamics that can serve as autonomous programmable quantum computers \cite{Ergodic}. Therefore, it is not unexpected that problems related to spectra and eigenvectors of self-adjoint operators lead to computationally hard problems. One could think that many of such problems could be solved efficiently on a quantum computer.  However, results proving that questions pertaining to the minimal eigenvalue of Hamiltonians are QMA-complete 
\cite{KitaevShen,Kempe2local,Oliveira} 
demonstrate that it is unlikely to find efficient algorithms for this problem.

The situation changes dramatically when we do not aim at deciding whether some Hamiltonian $H$ has an eigenvalue below a certain bound but only whether a given state $|\psi\rangle$ has a considerable component in the eigenspace corresponding to a particular eigenvalue of $H$.  This problem can be answered by (1) applying $H$-measurements to $|\psi\rangle$ several times and (2) statistically evaluating the obtained samples.  It has been shown in \cite{WZ:06} that measurements of so-called $k$-local operators\footnote{An
operator on $n$ qubits is 
called $k$-local if it can be decomposed into 
a sum of terms that act on at most $k$ qubits \cite{KitaevShen})}, applied to a basis state, solve all problems in BQP. This proves that some class of
problems concerning the spectral measure of $k$-local self-adjoint operators associated with a given state, 
characterize the class of problems that can be solved efficiently on a quantum computer.  Unfortunately,  the
requirement of $k$-locality restricts the applicability of these results since it is not clear where $k$-local matrices occur apart from in the study of quantum systems.  For this reason we have  constructed a problem with sparse matrices that does not require such a $k$-local structure and show that a very natural problem, namely the computation of diagonal entries of their powers, characterizes the complexity class BQP.

The paper is organized as follows.  In Section~\ref{DefDEE} we define formally the problem of estimating diagonal entries. In Section~\ref{inBQP} we prove that this problem can be solved efficiently on a quantum computer. To this end, we use the quantum phase estimation algorithm to implement a measurement of the
observable defined by the sparse matrix. To do this it is necessary that the time evolution defined by the sparse matrix can be implement efficiently.  Since the diagonal entries of the $m$th powers are the $m$th statistical moments of the spectral measure, we can estimate them after  polynomially many measurements provided that the accuracy is sufficiently high.  In Section~\ref{BQPhard} we show that diagonal entry estimation encompasses BQP.  The proof relies on an encoding of the quantum circuit which solves the considered computational problem into a sparse self-adjoint matrix such that the spectral measure (and hence
an appropriately chosen statistical moment) corresponding to the initial state depends on the solution.  In Section~\ref{General} we generalize this result to matrices with entries $-1$, $0$, and $1$. The idea is that the gates, which are encoded into the constructed Hamiltonian are not required to be unitary, even though
the circuit that then realizes the corresponding measurement is certainly unitary. This fact could be interesting in its own right.  For example, it could be possible that there are even more general
ways of simulating non-unitary circuits by encoding them into self-adjoint operators. In this context, it would be interesting to clarify the relation to other measurement based schemes of computation
\cite{RB00,browne-2006-,childs-2005-71}.

\section{Definition of diagonal entry estimation}

\label{DefDEE}
Before we define the decision problem ``diagonal entry estimation'' we
 introduce the notion of sparse matrices and spectral measures.
Here  we call an $N\times N$ matrix $A$ sparse if
it has no more than $s=\polylog(N)$ non-zero entries in each row and
there is an efficiently computable function $f$ that specifies for a
given row the non-zero entries and their positions  (compare 
\cite{ATS,ChildsDiss,BACS:06}).

Let $A$ be a self-adjoint matrix of size $N \times N$ and
\[
A=\sum_{\lambda} \lambda\, Q_\lambda
\]
be the spectral decomposition of $A$.  Let $|\psi\rangle$ be some
normalized vector of size $N$.  The spectral measure induced by $A$
and $|\psi\rangle$ is a probability distribution on the spectrum of $A$
such that the eigenvalue $\lambda$ occurs with probability
$\|Q_\lambda|\psi\rangle\|^2$.  In the sequel we will repeatedly make
use of the following observation. The expectation value of $A^m$ in
the state $|\psi\rangle$ is given by 
\[
\langle \psi |A^m|\psi\rangle=
\sum_\lambda \lambda^m \langle \psi|Q_\lambda|\psi\rangle\,,
\]
i.e., by the $m$th statistical moment of the spectral measure. 

In \cite{WZ:06}, eigenvalue sampling
is defined to be a quantum process that allows us to sample from a
probability distribution that coincides with the spectral measure induced
by $A$ and $|\psi\rangle$.  
Throughout the paper we
refer to such a
procedure as measuring the observable $A$ in the state $|\psi\rangle$.
We now state the considered problem in a formal way.

\begin{Definition}[Diagonal Entry Estimation]\label{DEE}${}$\\
Given a sparse real symmetric matrix $A$ of size $N$, an integer
$j\in\{1,\ldots,N\}$, and a positive integer $m=\polylog(N)$, estimate
the diagonal entry $(A^m)_{jj}$ in the following sense:

Decide if either 
\[
(A^m)_{jj} \geq g + \epsilon\, b^m
\]
or 
\[
(A^m)_{jj} \leq g - \epsilon\, b^m\,,
\]
for given $g\in [-b^m,b^m]$ and $\epsilon=1/\polylog(N)$, where $b$ is
an a priori known upper bound on the operator norm of $A$.
\end{Definition}

\medskip
Problems of this kind arise, for example, in graph theory. Let $A$ be
the adjacency matrix of a graph with $N$ vertices and degree
bounded from above by $s$.  Then the
diagonal entry $(A^m)_{jj}$ of the $m$th power of $A$ is equal to the
number of paths of length $m$ that start and end at the vertex $j$.

It is important to note that the scale on which the estimation has
reasonable precision is given by $b^m$.  If the
a priori known bound on the norm is, for instance,  $b':=2b$ instead of $b$, 
then the
accuracy is changed by the exponential factor $2^m$. 
Our results
show that quantum computation outperforms classical computation in
estimating the diagonal entries (provided that BQP$\neq$ BPP). But one
has to be very careful on which scale this result remains true.

\section{Diagonal entry estimation is in BQP}\label{inBQP}
To show that diagonal entry estimation is in BQP we briefly recall
the formal definition of this complexity class \cite{KitaevShen}.

\begin{Definition}[The class BQP]\label{BQP}${}$\\
A language $L$ is in BQP if and only if there is a
uniformly generated family of quantum circuits $Y_r$ acting on
$\poly(r)$ qubits that decide if a string ${\bf x}$ of length $r$ is
contained in $L$ in the following sense:
\begin{equation}\label{Schalt}
Y_r|{\bf x},{\bf 0}\rangle = 
\alpha_{{\bf x},0} |0\rangle\otimes |\psi_{{\bf x},0}\rangle +
\alpha_{{\bf x},1} |1\rangle \otimes |\psi_{{\bf x},1}\rangle
\end{equation}
such that
\begin{enumerate}
\item $|\alpha_{{\bf x},1}|^2 \ge 2/3$ if ${\bf x}\in L$ and
\item $|\alpha_{{\bf x},1}|^2 \le 1/3$ if ${\bf x}\not\in L$\,.
\end{enumerate}
Equation~(\ref{Schalt}) has to be read as follows. The input string
${\bf x}$ determines the first $r$ bits. Furthermore,  $l$ additional
ancilla bits are
initialized to $0$. After $Y_r$ has been  applied we interpret
the first qubit as the relevant 
output and the remaining $r+l-1$ output values  
are irrelevant. 
The size of the ancilla register is polynomial in $r$.
\end{Definition}

We now describe how to construct a circuit that solves diagonal entry estimation.  Without loss of generality we may assume $b=1$ and rescale the measurement results later.  The main idea is as follows.  We measure the observable $A$ in the state $|j\rangle$ and raise the outcome value to the $m$th power. The average value of the obtained values over large sampling converges to the desired entry.  The measurement is done by (a) considering $A$ as a Hamiltonian of a quantum system and simulating the corresponding dynamics $U_t=\exp(-i A t)$ and (b) applying the phase estimation algorithm to $U_t$.  The proof that this works follows from a careful analysis of possible error sources.  These are (1) errors due to the statistical nature of the phase estimation algorithm, (2) statistical errors due to estimation of the expectation value from the empirical mean, and (3) errors caused due to the imperfect simulation of the Hamiltonian time evolution.  We show that all these errors can be made sufficiently small with polynomial resources only.

(1) We embed $A$ into the Hilbert space of $n$ qubits, where $n=\lceil \log_2 N \rceil$.  Let us first assume that the unitary 
$U:=\exp(i A)$ can be implemented exactly.  We apply the phase estimation procedure which works as follows \cite{NC}.  
We start by adding $p$ ancillas to the qubits on which $U$ acts.  The idea is to control the implementation of the $2^l$th power of $U$ by the $l$th control qubit.  More precisely, we have the controlled gates
\[
W_l := 
|0 \rangle\langle 0|^{(l)} \otimes {\bf 1} + 
|1 \rangle\langle 1|^{(l)} \otimes U^{2^l}\,,
\]
where the superscript ${(l)}$ indicates that the projectors $|0\rangle\langle 0|$ and $|1 \rangle\langle 1|$ act on the $l$th control qubit, respectively.  Note that the decomposition of $W_1$ into elementary gates is obtained by replacing each elementary gate in the circuit implementing $U$ with a corresponding controlled
gate. Similarly, $W_l$ is realized by applying the quantum circuit implementing the corresponding controlled $U$-gate $2^l$ times. 
Set $W:=W_1 W_2 \cdots W_p$.  
%Clearly, this unitary can be implemented with $O(2^p)$ elementary gates.
The phase estimation circuit consists of applying Hadamard gates on all control qubits, the circuit $W$, and the inverse Fourier transform on the control qubits.  
The desired value $a$ is obtained by measuring the control qubits in the computational basis.
Let $|\psi\>$ be an arbitrary eigenvector of $U$ with unknown eigenvalue $e^{i2\pi \varphi}$ for some phase $\varphi\in[0,1)$.  In order to achieve
that
the phase estimation algorithm outputs a random 
value $a\in\{0,\ldots,2^p-1\}$ such that
\begin{equation}\label{ErrorB}
\mathrm{Pr}(|\varphi - a/2^p|<\eta)>1-\theta\,,
\end{equation}
for some $\theta,\eta >0$ 
it is sufficient \cite{NC} to set
\[
p:=
\lceil \log(1/\eta)\rceil + \lceil \log\big(2+(1/(2\theta)\big) \rceil\,.
\]
%The running time of the phase estimation is $O(2^p)$. 

Let $|\psi\>$ be an eigenvector of $A$ with unknown 
eigenvalue $\lambda\in [-1,1]$.  
%It gives rise to the phase $\varphi=\lambda/2\pi$ of the unitary $U=\exp(2\pi i (A/2\pi))$. 
In order to determine $\lambda$ approximately using the outcome
$a$ in a phase estimation with $U=\exp(iA)$ 
we proceed as follows. First, we have to take into account that
$\varphi >1/2$ corresponds to negative values $\lambda$. Second,
we have to consider that the scaling differs by the factor $2\pi$.
Finally, we may use the additional information that not all $\lambda$
in $[-\pi,\pi)$ are possible, but only those in $[-1,1]$. 
All outputs that would actually correspond to eigenvalues $\lambda$ 
in  $[1,\pi]$ and $[-\pi,-1)$ are therefore interpreted
as $\pm 1$, repectively.
Therefore, we compute values $z$ from the output $a$ by
\[
z:=
\left\{
\begin{array}{lll}
a (2\pi/2^p)            & \mbox{for} & 0              \le a < 
2^p/(2\pi)     \\
1                       & \mbox{for} & 2^p/(2\pi)     \le a < 
2^p/2          \\
-1                      & \mbox{for} & 2^p/2          \le a < 
2^p-2^p/(2\pi) \\
a (2\pi/2^p)-2\pi     & \mbox{for} & 2^p-2^p/(2\pi) \le a < 2^p
\end{array}
\right.
\]
This defines the random variable $Z$ whose values $z$ are  
approximations for $\lambda$ that satisfy the following error bound:
\[
\mathrm{Pr}
\big( | \lambda - Z | < 2\pi\eta \big) > 1 - \theta\,.
\]
This bound follows from ineq.~(\ref{ErrorB}) by appropriate rescaling
(note that our reinterpretation of values in $[-\pi,-1]$ and
$[1,\pi)$ explained above can only decrease the error unless
it was already greater than $\pi-1$). 
Consequently, we have for every eigenstate $|\psi_i\rangle$ of $A$ 
with eigenvalue $\lambda_i$ the statement
\begin{equation}
\label{thetain} |E_{|\psi_i\rangle}(Z^m)-\lambda^m_i|
\leq
2\,\theta  + 2\pi m\, \eta\,,
\end{equation}
where $E_{|\psi_i\rangle}(Z^m)$ denotes the expectation value of $Z^m$ 
in the state $|\psi_i\rangle$. The first term on the right hand side 
corresponds to the unlikely case that 
the measurement outcome deviates by more 
than $2\pi \eta$ from the true value. Since we do not have outcomes $z$ 
smaller than $-1$ or greater than $1$ the maximal error is at most $2$. 
This leads to the error term $2\theta$. The second term corresponds to 
the case $|\lambda_i-z|\leq 2\pi \eta$, which implies $|\lambda^m_i-z^m| 
\leq (2\pi \eta) m$ because $\lambda_i,z\in [-1,+1]$.

We make the error in eq.~(\ref{thetain}) smaller than $\epsilon/3$ by choosing the parameters $\theta$ and $\eta$ such that $\theta<\epsilon/12$ and $\eta<\epsilon/(12\,\pi\, m)$.  The number of control qubits can be chosen to be 
\begin{equation}\label{pDef}
p:=2\lceil\log((48\, m)/\epsilon)\rceil\,.
\end{equation}
This is sufficient since 
\begin{equation}\label{eq:upperBound_p}
\lceil \log(1/\eta)\rceil + \lceil \log\big(2+(1/(2\theta)\big) <
%\rceil\lceil \log(1/\eta)\rceil + \lceil \log\big(2+(1/(2\theta)\big) \rceil < 
%\lceil \log\big((12\pi m)/\epsilon\big)\rceil + 
%\lceil \log\big(2+6/\epsilon\big) \rceil < 
2\lceil \log\big( (48\, m)/ \epsilon \big) \rceil\,.
\end{equation}

We decompose $|j\rangle$ into $U$-eigenstates
\[
|j\rangle=\sum_i \beta_i |\psi_i\rangle\,.
\]
and obtain the statement 
\[
E_{|j\rangle}(Z^m)=\sum_i |\beta_i|^2 \,E_{|\psi_i\rangle}(Z^m)
\]
by linearity arguments and 
\[
(A^m)_{jj}=\sum_i |\beta_i|^2 \lambda_i^m\,.
\]
Using the triangle inequality and the fact that the right hand side of
ineq.~(\ref{thetain}) is smaller than $\epsilon/3$ for each $i$ we obtain 
\begin{equation}\label{Eps}
|E_{|j\rangle}(Z^m)-(A^m)_{jj}| < \epsilon/3\,.
\end{equation}

(2) Now we sample the measurement $k$ times in order to estimate the expectation value $E_{|j\rangle}(Z^m)$.
Since we will later also consider the simulation error we want now  estimate $(A^m)_{jj}$ up to an error of $2\epsilon/3$.  To achieve this, it is sufficient to estimate $E_{|j\rangle}(Z^m)$ up to an error of $\epsilon/3$.  

Let $\overline{Z^m}$ denote the average value of the random variable $Z^m$ after sampling $k$ times.  Since the values of $Z^m$ are between $-1$ and $1$ we can give an upper bound for the probability to obtain an average being not $\epsilon/3$-close to the expectation value. By Hoeffding's inequality \cite[Theorem~2]{Hoeffding} we get
\begin{eqnarray*}
{\rm Pr}\Big(|\overline{Z^m}-E_{|j\rangle}(Z^m)|
\, \geq\, 
\frac{\epsilon}{3} \, \Big) 
& \leq & 
2 \exp\Big(\frac{-\epsilon^2}{18}\, k\Big)\,.
\end{eqnarray*}

In summary, we have shown for $b=1$ that we can distinguish between the two cases in Definition~\ref{DEE} with exponentially small error probability.  For $b \neq 1$ we have to rescale the inaccuracy of the estimation by $b^m$.  The whole procedure including repeated measurements and averaging
can certainly be performed by a single quantum circuit $Y_r$ in the sense of Definition~\ref{BQP}.  

(3) We now take into account that $U=\exp(iA)$ cannot be implemented exactly.  It is known that 
 the dynamics generated by $A$ can be simulated efficiently if $A$ is sparse \cite{ATS,ChildsDiss,BACS:06}.  More precisely, for each $t$  we can construct a circuit $V$ which is $\delta$-close to $U_t=\exp(-i A t)$ with respect to the operator norm  such that the required number of gates increases only polynomially in the parameters $n,\,t,$ and $1/\delta $. We analyze the error resulting from using $V$ instead of $U$, where $\|V-U\|\leq \delta$.  

The phase estimation contains $2^{p+1}-1$ copies of the controlled-$V$ gate.  Therefore the circuit $F_V$ implementing the phase estimating procedure with $V$ instead of $U$ deviates from $F_U$ by at most $2^{p+1}\,\delta$ with respect to the operator norm, that is, $\|F_U-F_V\|\le 2^{p+1}\,\delta$.  

Let $q$ and $\tilde{q}$ denote the probability distributions of outcomes when measuring the control register after the phase estimation procedure has been implemented with $V$ and $U$, respectively.  The $l_1$- distance between $q$ and $\tilde{q}$ is then defined by 
\[
\|q-\tilde{q}\|_1:=
\sum_{a\in\{0,\ldots,2^p-1\}} |q(a) - \tilde{q}(a)|\,
\]
where $q(a)$, respective $\tilde{q}(a)$, denote the probabilities of obtaining the outcome $a$.  To upper bound $\|q-\tilde{q}\|_1$ we define a function $s$ by $s(a):=1$ if $q(a)>\tilde{a}$ and $s(a):=-1$ otherwise.
Let $Q$ be the observable defined by measuring the ancillas and applying $s$ to the outcome $a$.  Then we can write $\|q-\tilde{q}\|_1$ as a difference of expectation values: 
\[
\langle \psi |F^\dagger_U QF_U- F^\dagger_V QF_V|\psi\rangle
\leq 2 \|F_U-F_V\|\,\|Q\|\leq 2^{p+2}\,\delta\,.
\]
This implies that the corresponding expectation values of  $Z^m$ can differ by at most $2^{p+2} \,\delta$ because $Z$ takes only values in the interval $[-1,1]$.  We choose the simulation accuracy such that $\delta=\epsilon/(3\cdot 2^{p+2})$ and obtain an additional error term of at most $\epsilon/3$ in ineq.~(\ref{Eps}).  Using 
%the upper bound in eq.~(\ref{eq:upperBound_p}) 
that we have chosen $p$ as in eq.~(\ref{pDef})
we obtain that $\delta\in O(\epsilon^3 m^2)$.

Putting everything together we obtain a total error of at most $\epsilon$.  Furthermore, this can be achieved by using time and space resources which are polynomial in $n$, $m$, and $1/\epsilon$.  This completes the proof that diagonal entry estimation is in BQP.

%hence  
%$\delta$ is inverse polynomial, too. This ensures that the 
%running time of the circuit, including simulation and
%phase estimation procedures, is polynomial. 
%The space resources are polynomial, anyway, since  only a logarithmic 
%number of ancillas is needed in addition to the $n$ qubits where
%$A$ acts on. Since we have furthermore shown that the 
%number of required runs for the sampling procedure 
%is only logarithmic with the inverse of $\epsilon$ we have shown that
%the problem is in BQP.

It should be mentioned that off-diagonal entries $(A^m)_{ij}$ can also be estimated efficiently on the same scale using superpositions $|i\rangle \pm |j\rangle$ since the values $\langle i |A^m |j\rangle$ can be expressed in terms of differences of the statistical moments of the spectral measure induced by those states. The scale on which the estimation can be done efficiently is then also given by $\epsilon \,b^m$ with an appropriately modified $\epsilon$ which is still inverse polynomial in $n$. However, since BQP hardness requires only  diagonal entries we have focused our attention on the latter.

\section{Diagonal entry estimation is BPQ-hard}
\label{BQPhard}

Now we assume that we are given  a quantum circuit $Y_r$ that
is able to decide whether a string ${\bf x}$ is in the given language $L$ 
in the sense of Definition~\ref{BQP}. Using
$Y_r$ 
we construct a self-adjoint operator $A$ such that
the
corresponding spectral measure induced by an appropriate initial state depends on whether ${\bf x}$ is in $L$ or not.  
Note that the proofs for QMA-completeness of eigenvalue problems
for  Hamiltonians have already used the idea to 
construct a self-adjoint operator whose spectral properties
encode a given quantum circuit \cite{KitaevShen,Kempe2local,Oliveira}.
In these constructions, the {\it existence} of eigenvalues
of a given Hamiltonian depends on whether or not an input state
exists that is accepted by a certain circuit. In BQP, the problem is
only to decide whether a 
{\it given} state is accepted and not whether such a state exists.
Likewise, the problem is not to decide whether an eigenvalue
of the constructed observable {\it exists} which lies 
in a certain interval. Instead, it refers only to the
the spectral measure induced by  a {\it given} state. This difference 
 changes
the complexity from QMA to BQP. 

The idea for our construction is therefore rather 
based on \cite{WZ:06} which shows the BQP-hardness of approximate $k$-local measurements.  This result relied on the ideas in \cite{PSPACE} where the PSPACE-hardness of so-called exact $k$-local measurements was proved.

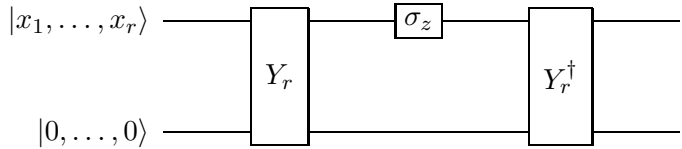
\begin{figure}
\centerline{
\Qcircuit @C=3em @R=2.8em {
%\lstick{|1\rangle} &  \qw  & \gate{-1}  &\qw &\qw \\
\lstick{|x_1,\dots,x_r\rangle}&    \multigate{1}{Y_r} &\gate{\sigma_z}
&\multigate{1}{Y_r^\dagger} &\qw  \\
\lstick{|0,\dots,0\rangle}&  \ghost{Y_r} &\qw & \ghost{Y^\dagger_r} &\qw
}
}
\vspace{0.5cm}
\caption{\label{Circ}{\small Circuit $U$ constructed from the original circuit
$Y_r$. Whenever the answer of the BQP problem is no, the output state
of $U$ is close to the input sate
$|{\bf x},{\bf 0}\rangle \equiv |x_1,\dots,x_n,0,\dots,0\rangle$. 
Otherwise, the state
$|{\bf x},{\bf 0}\rangle$ is only restored after applying $U$ twice. 
}}
\end{figure}

However, our description below will only at one point refer to these results since the observable we construct here is only required to be sparse, 
in contrast to the $k$-locality assumed in
\cite{WZ:06,PSPACE}.  
In some analogy to \cite{IdentityQMA,WZ:06} we construct a circuit $U$
that is obtained from $Y_r$ as follows: First apply the  circuit $Y_r$. Apply then a $\sigma_z$-gate.
Implement then $Y^{\dagger}_r$. The resulting circuit $U$ is shown in Fig.~\ref{Circ}.  We  denote the dimension of the Hilbert space $U$ acts on by $\tilde{N}$.

Let $U$ be generated by a concatenation of the
$M$ elementary gates $U_0,\dots,U_{M-1}$. We assume furthermore 
that $M$ is odd, which
is automatically satisfied if we decompose 
$Y_r^\dagger$ in analogy to $Y_r$ and
implement a $\sigma_z$-gate between $Y_r$ and $Y_r^\dagger$.  
We define the unitary
\begin{equation}\label{Ul}
W:=\sum_{l=0}^{M-1} |l+1\rangle\langle  l| \otimes U_l\,,
\end{equation}
acting on $\C^M\otimes \C^{\tilde{N}}$. 
Here the $+$ sign in the index has always to be read modulo $M$. 
We obtain
\[
W^M=\sum_{l=0}^{M-1} 
|l\rangle \langle l|\otimes U_{l+M} \, \cdots U_{l+1} \,  U_{l}\,.
\] 
Due to $U^2={\bf 1}$ we have $(W^M)^2={\bf 1}$.
Thus, $W^M$ can  only have the eigenvalues
$\pm 1$. This defines a decomposition of 
the space $\C^M\otimes \C^{\tilde{N}}$ into
symmetric and antisymmetric $W$-invariant 
subspaces $\cS^+$ and $\cS^-$, respectively with projections
\[
Q^\pm:=\frac{1}{2}({\bf 1} \pm W^M)\,.
\]
In the following we use the definition
\[
|s_{\bf x}\rangle:=|0\rangle \otimes |{\bf x},{\bf 0}\rangle
\] 
for the initial state and restrict the attention to the span of the orbit
\begin{equation}
\label{Orbit}
\Big\{\,W^l|s_{\bf x}\rangle  \,\Big\} \,\,\,\,\,\,\hbox{ with } l\in \N\,.
\end{equation}
Moreover, we use the abbreviations $\alpha_0=\alpha_{{\bf x},0}$ and
$\alpha_1=\alpha_{{\bf x},1}$.  We consider first the two extreme cases $|\alpha_1|=0$ and $|\alpha_1|=1$. 
If $|\alpha_1|=0$ the orbit (\ref{Orbit})  is $M$-periodic and the action of $W$ is isomorphic to the action of a cyclic shift in $M$ dimensions, i.e., the mapping $|l\rangle \mapsto |(l+1) \mod M\rangle$, where 
$|l\rangle$ corresponds to $W^l|s_{\bf x}\rangle$ with $l=0,1,\dots,M-1$.  

If $|\alpha_1|=1$ the action of $W$ corresponds to a cyclic shift with an additional phase $-1$, i.e., the mapping $|l\rangle \mapsto |l+1\rangle$ for $l=0,1,\dots,M-2$ and $|M-1\rangle \mapsto -|0\rangle$. In the first case, the state $|s_{\bf x}\rangle$ induces a spectral measure $R^{(0)}$ being the uniform distribution on the $M$th roots of unity, i.e., the values $\exp(-i\pi\, 2l/M)$ for $l=0,\dots,M-1$.  In the second case, $|s_{\bf x}\rangle$ induces the measure $R^{(1)}$ being the uniform distribution on the values $\exp(-i\pi\,(2l+1)/M)$ for $l=0,\ldots,M-1$. We observe that $R^{(1)}$ and $R^{(0)}$ coincide up to a reflection of the real axis in the complex plane. 

In the general case, the orbit defines an $2M$-dimensional space whose orthonormal basis vectors
are obtained by renormalizing the vectors 
\[
W^l Q^+|s_{\bf x}\rangle \quad \hbox{ and } \quad
W^l Q^-|s_{\bf x}\rangle \quad \hbox{ with } 
l=0,1,\dots,M-1\,.
\]
We obtain then a convex sum of $R^{(0)}$ and $R^{(1)}$ as spectral measures induced by $W$ and 
$|s_{\bf x}\rangle$. The following calculation shows that $|\alpha_0|^2$ and $|\alpha_1|^2$ define the corresponding weights:
\begin{eqnarray*}
\langle s_{\bf x} |Q^+|s_{\bf x}\rangle &=&
\frac{1}{2}\langle s_{\bf x} |{\bf 1} +W^M|s_{\bf x}\rangle\\
&=&\frac{1}{2} \langle {\bf x},{\bf 0}| {\bf 1} +U|{\bf 0},{\bf x} \rangle\\
&=& \frac{1}{2} (1+ \langle {\bf x},{\bf 0}|Y_r^\dagger 
\sigma_z Y_r|{\bf 0},{\bf x} )\rangle\\
&=&  |\alpha_0|^2\,,
\end{eqnarray*}  
where the last equality follows easily by replacing
$Y_r|{\bf x},{\bf 0}\rangle$  and its adjoint with
the expression in eq.~(\ref{Schalt}) and its adjoint.
Thus, we obtain the spectral measure
\[
R:=|\alpha_0|^2 R^{(0)}+ |\alpha_1|^2 R^{(1)}\,.
\]
We define the self-adjoint operator
\[
A:=\frac{1}{2}(W+W^\dagger)\,.
\] 
The support of the spectral measure corresponding to $A$ is directly given by
the real part of the support of $R$. 
To obtain the corresponding probabilities one has to take
into account that in many cases two different eigenvalues of $W$ 
lead to the same
eigenvalue of $A$. 

To calculate the distribution of outcomes for $A$-measurements
we observe that $R^{(0)}$ leads therefore to a distribution
$P^{(0)}$
 on
the $(M-1)/2$ eigenvalues 
\[
\lambda^{(0)}_l= \cos\frac{ 2\pi l}{M}
\,\,\,\,\, 
\hbox{ for } l=0,\dots,(M-1)/2
\]
with  probabilities $P^{(0)}_1:=1/M$ and $P^{(0)}_l:=2/M$ for
$l > 1$. Likewise,  $R^{(1)}$ leads to a distribution $P^{(1)}$ on
the $(M-1)/2$ values 
\[
\lambda^{(1)}_l=\cos \frac{\pi (2l+1)}{M}\,\,\,\,\, \hbox{ for } 
l=0,\dots,(M-1)/2
\]
with  probabilities $P^{(1)}_{(M-1)/2}=1/M$ and $P^{(1)}_l=2/M$ for $l < (M-1)/2$. As it was true for $R^{(0)}$ and $R^{(1)}$, the measures $P^{(0)}$ and $P^{(1)}$ coincide up to a reflection.  

We now set $|j\rangle:=|s_{\bf x}\rangle$, i.e.,
the input state is considered as the $j$th basis vector of
$\C^M\otimes \C^{\tilde{N}}$.  
Then the diagonal entry $(A^m)_{jj}$ coincides
with the $m$th moment of the spectral measure:
\[
(A^m)_{jj}= \langle j|A^m|j\rangle=\sum_\lambda \lambda^m P(\lambda)\,,
\]
where $\lambda$ runs over all eigenvalues of the restriction of $A$ 
to the smallest $A$-invariant subspace containing $|j\rangle$, 
and $P(\lambda)$  denotes its probability according to the 
spectral measure corresponding to $A$.  Since the latter is a convex sum of
$P^{(0)}$ and $P^{(1)}$ we may write $(A^m)_{jj}$ as the convex sum
\begin{eqnarray}\nonumber
(A^m)_{jj}&= &(1-|\alpha_1|^2)\sum_l \Big(\lambda_l^{(0)}\Big)^m
P_l^{(0)}
+ |\alpha_1|^2\sum_l \Big(\lambda_l^{(1)}\Big)^m
P_l^{(1)}\\
&=:&(1-|\alpha_1|^2)\, E_0 
+ |\alpha_1|^2\, E_1
\,.\label{Amjj}
\end{eqnarray}
The values $E_0$ and $E_1$ can be considered as the $m$th statistical
moments of random variables on $[-1,1]$ whose distributions
are given by $P^{(0)}$ and $P^{(1)}$, respectively.

In order to see how the value $(A^m)_{jj}$ changes with $|\alpha_1|$
we observe
\[
E_0=\sum_{l=0}^{(M-1)/2} \Big(\lambda_l^{(0)}\Big)^m
P_l^{(0)} \geq P_0^{(0)}+\Big(\lambda_{(M-1)/2} ^{(0)} \Big)^m 
=\frac{1}{M}+\Big(\lambda_{(M-1)/2} ^{(0)} \Big)^m \,.
\]
Here we have used $\lambda_0^{(0)}=1$ and 
that the eigenvalues are numbered in a decreasing
order. Thus, $\lambda_{(M-1)/2}$ is the smallest one. 
Due to the reflection symmetry of the measures 
we
have $E_1=-E_0$. Now we choose $m$ sufficiently large 
such that the term $(\lambda^{(0)}_{(M-1)/2})^m$ is negligible
compared to $1/M$ since we have then $E_0-E_1 \approx 2/M$ which 
is a sufficient difference for our purpose. 
 
In order to achieve this we
set $m:=M^3$.
We have 
\[
\lambda_{(M-1)/2}^{(0)}=-\cos(\pi/M)> 
-1+\frac{\pi^2 }{2 M^2} -\frac{\pi^4}{4!\,M^4}
>
-1+\frac{\pi^2}{4\,M^2}\,, 
\]
where the last inequality holds for sufficiently large $M$. 
Due to 
\[
\lim_{M\to \infty} (1-\frac{\pi^2}{4\,M^2})^{M^2}=e^{\frac{-\pi^2}{4}}
\] 
we conclude that 
\[
(\cos(\pi/M))^{M^3}< (e^{-\frac{\pi^2}{4}})^M \,,
\]
and hence
\begin{equation}\label{E0}
E_0> \frac{1}{M}- (e^{- \frac{\pi^2}{4}})^M >\frac{3}{4M} \,,
\end{equation}
where we have, again, assumed $M$ to be sufficiently large.
To see  how $(A^m)_{jj}$ changes with $|\alpha_1|$ 
we recall
\[
(A^m)_{jj}= (1-|\alpha_1|^2)E_0
+
|\alpha_1|^2E_1=(1-2|\alpha_1|^2) \,E_0\,,
\]
by eq.~(\ref{Amjj}) and the reflection symmetry. 
Using the worst cases $|\alpha_1|^2=1/3$ for $x\not\in L$ and 
$|\alpha_1|^2=2/3$ for $x\in L$ we 
obtain 
\[
(A^m)_{jj}=\frac{1}{3} E_0 
\]
and
\[
(A^m)_{jj}=-\frac{1}{3} E_0\,.
\]
Using $E_0 > 3/(4M)$ from ineq.~(\ref{E0}) we obtain 
\[
(A^m)_{jj} > \frac{1}{4M}\,, 
\]
if the answer is no and
\[
(A^m)_{jj} < -\frac{1}{4M}
\]
otherwise.
Setting  $g:=0$ (see  Definition~\ref{DEE}) 
we may define $\epsilon:=1/(4\,M)$. Then
the diagonal entry is greater than $g+\epsilon$ if $x\not\in L$ and
smaller than  $g-\epsilon$ otherwise.
The construction of $A$ as the real part of a unitary 
ensures that $\|A\|\leq 1=:b$.  
This shows that we can find an inverse
polynomial accuracy $\epsilon$ such that an estimation of the
diagonal
entry up to an error $\epsilon\, b^m$ 
allows to check whether $x$ is in
$L$.

\section{Generalization to matrices with entries $0,\pm 1$}

\label{General}

So far we have allowed general real-valued matrix entries. We may
strengthen the result of the preceding section in the sense that diagonal entry estimation remains BQP-hard
if we allow matrix entries to be only $0,\pm 1$.  It is known that Toffoli gates and Hadamard gates form together a universal set for quantum computation \cite{aharonov-2003-}.  We may thus replace the whole sequence $U_1,\dots,U_N$ of gates used in the definition of $W$ (see eq.~(\ref{Ul})) by a set of gates that consist only of Toffoli and Hadamard gates. Let $T$  and $H$, denote the set of Toffoli gates and the set of Hadamard gates on $\C^n$, respectively.  We  modify then the universal set and consider $T_{left} \,\cup T_{right}\, \cup H$, where we have defined $T_{left} :=TH$ and $T_{right}:=HT$. In words, $T_{left}$ is, for instance,  the set of gates that are obtained by applying an arbitrary Toffoli-gate followed by a Hadamard gate on an arbitrary qubit.  The Toffoli gates are permutation matrices, 
whereas the Hadamard matrices have only entries $\pm 1/\sqrt{2}$. 
Thus, the
gates in $T_{right}$ and $T_{left}$ 
have only entries $\pm 1/\sqrt{2}$, too.
Therefore, all gates in $T_{left}\,\cup T_{right}\,\cup H$ have  only entries
 $\pm 1/\sqrt{2}$. Hence, the matrix $A$ would only consist of such entries when using only gates that are taken from our modified set of gates. By rescaling with $\sqrt{2}$ we obtain a matrix $A$ with
entries $0,\pm 1$. The rescaling is clearly irrelevant for
the diagonal entry estimation problem since we have now 
spectral values within the interval $[-\sqrt{2},\sqrt{2}]$ and the  
accuracy required  by Definition \ref{DEE} changes by  the
factor
$(\sqrt{2})^m$ accordingly.

\section{Conclusions}

We have shown that the estimation of diagonal entries of powers of
symmetric sparse matrices is BQP complete when the demanded accuracy scales
appropriately with the powers of the operator norm. 

The quantum algorithm proposed here for solving this problem 
uses the fact that measurements of the corresponding observable 
allow to obtain enough information on the probability measure
defined by the eigenvector decomposition of the considered basis
state. Given the assumption that $BQP\neq BPP$, i.e., that a quantum
computer is more powerful than a classical computer, the
required information on the spectral measure cannot be obtained by 
any efficient classical algorithm.  This is remarkable since the determination of spectral measures is a problem whose relevance is not restricted to applications in quantum theory only.

\section*{Acknowledgments}
P.W. was supported by the Army Research Office under Grant No. W911NF-05-1-0294 and by the National Science Foundation under Grant No. PHY-456720.  This work was begun during D.J.'s visit at the Institute for Quantum Information at the California Institute of Technology.  The hospitality of the IQI members is gratefully acknowledged.  Moreover, we would like to thank Shengyu Zhang and Andrew Childs for helpful discussions.

%\bibliographystyle{unsrt}

%\bibliography{/home/janzing/literatur,/home/janzing/Caltechliteratur}

\end{document}